\documentclass[%
 reprint,
 showpacs,
 amsmath,amssymb,
 aps,
 pra,
]{revtex4-1}

\usepackage[ascii]{inputenc}
\usepackage[T1]{fontenc}
\usepackage[english]{babel}
\usepackage{graphicx}
\usepackage{microtype}
\usepackage[pdftex,hidelinks]{hyperref}
\hypersetup{%
  pdftitle={Quantum master equation with balanced gain and loss},%
  pdfauthor={Dennis Dast, Daniel Haag, Holger Cartarius, G\"unter Wunner}%
}
  
\DeclareMathOperator{\imag}{Im}
\DeclareMathOperator{\real}{Re}
\DeclareMathOperator{\trace}{tr}
\newcommand{\PT}{\mathcal{PT}}
\newcommand{\mbf}{\mathbf}
\newcommand{\mrm}{\mathrm}
\newcommand{\rmi}{\mathrm{i}}
\newcommand{\rmd}{\mathrm{d}}

\newcommand{\bra}[1]{\langle{#1}|}
\newcommand{\ket}[1]{|{#1}\rangle}

\newcommand{\braopket}[3]{\langle{#1}|{#2}|{#3}\rangle}
\newcommand{\mean}[1]{\langle{#1}\rangle}

\begin{document}

\title{Quantum master equation with balanced gain and loss}

\author{Dennis Dast}
\email[]{dennis.dast@itp1.uni-stuttgart.de}

\author{Daniel Haag}

\author{Holger Cartarius}

\author{G\"unter Wunner}

\affiliation{Institut f\"ur Theoretische Physik 1,
  Universit\"at Stuttgart, 70550 Stuttgart, Germany}

\date{\today}

\begin{abstract}
  We present a quantum master equation describing a Bose-Einstein condensate
  with particle loss on one lattice site and particle gain on the other lattice
  site whose mean-field limit is a non-Hermitian $\PT$-symmetric
  Gross-Pitaevskii equation.
  It is shown that the characteristic properties of $\PT$-symmetric systems,
  such as the existence of stationary states and the phase shift of pulses
  between two lattice sites, are also found in the many-particle system.
  Visualizing the dynamics on a Bloch sphere allows us to compare the complete
  dynamics of the master equation with that of the Gross-Pitaevskii equation.
  We find that even for a relatively small number of particles the dynamics are
  in excellent agreement and the master equation with balanced gain and loss is
  indeed an appropriate many-particle description of a $\PT$-symmetric
  Bose-Einstein condensate.
\end{abstract}


\maketitle

\section{Introduction}
\label{sec:introduction}
Since the seminal paper by Bender and Boettcher \cite{Bender98a} much progress
has been achieved formulating a consistent quantum theory in which the
requirement of Hermiticity is replaced by the weaker requirement of $\PT$
symmetry \cite{Bender99a, Bender07a} or pseudo-Hermiticity
\cite{Mostafazadeh02a, Mostafazadeh02b, Mostafazadeh02c}.
In addition complex $\PT$-symmetric potentials are used for an effective
description of quantum systems in contact with an environment.
Positive imaginary parts describe a source for the probability amplitude,
negative imaginary contributions lead to a sink.
In $\PT$-symmetric systems features can be observed that are not present in a
purely Hermitian quantum system.
They exhibit true stationary states in spite of an in- and outflux
of the probability amplitude \cite{Klaiman08a, Ruter10a, Guo09a, Peng14a,
Chong11a, Schindler11a, Bittner12a, Kreibich13a, Cartarius12b, Dast13b,
Graefe08b, Graefe10a, Haag14a}, the occurrence of exceptional points at which
two or more eigenstates coalesce \cite{Klaiman08a, Cartarius12b, Dast13b,
Graefe08b, Graefe10a, Haag14a}, complicated stability properties of the
stationary states, and a very rich dynamics \cite{Haag14a}.

The first experimental realization of a $\PT$-symmetric system succeeded in
optical waveguides \cite{Klaiman08a, Ruter10a, Guo09a, Peng14a}, and
theoretical proposals for various further systems exist \cite{Chong11a,
Schindler11a, Bittner12a, Kreibich13a}.
Although the concept of $\PT$ symmetry originates from quantum theory no
genuine $\PT$-symmetric quantum system has been realized so far.
This is, however, of great importance since the optical systems can only in
special cases correctly model effects of the Schr\"odinger equation.
An experimental realization in a genuine quantum system would provide a solid
basis for the theoretically investigated physical effects.

A $\PT$-symmetric quantum system which is potentially experimentally accessible
is a Bose-Einstein condensate in a double-well potential, in which particles
are removed from one well and injected into the other.
In both the idealized version of a double-$\delta$ potential
\cite{Cartarius12b} and in an spatially extended double well \cite{Dast13b} it
was shown that the system supports stationary solutions which are stable
with respect to small perturbations.

These investigations were done in the mean-field limit described by the
Gross-Pitaevskii equation, which is known to be accurate in the limit of
temperatures well below the critical temperature.
The gain and loss is modeled via an imaginary potential which is interpreted as
a coherent in- and outcoupling whose strength is proportional to the amount of
particles in the condensate.
The physical reasoning behind this proportionality is the Bose stimulation of
the incoupling, i.e.\ the transition rate is enhanced by a factor of $(N+1)$ if
there are already $N$ particles in the final state \cite{Hope96a, Miesner98a}.
Proposals for the experimental realization of such a complex $\PT$-symmetric
potential by embedding the system into a larger Hermitian transport structure
\cite{Kreibich13a} or via a coupling approach \cite{Single14a} were, again,
formulated in the mean-field limit.
In that limit all correlations are neglected, and, in addition, the condensate
is described as a pure state although for $\PT$-symmetric systems we are
especially interested in the coupling to the environment.

However, the only physical process describing a gain or loss for the wave
function of a Bose-Einstein condensate on the microscopic scale can be an
addition or removal of single particles.
Thus, there should exist a microscopic description.
It is the purpose of this article to demonstrate that this microscopic
description can be achieved.
Thus, it is possible to show that indeed in- and outcoupling processes for
single atoms exist, which are capable of explaining the origin of the complex
$\PT$-symmetric potentials in the mean-field limit.

On a microscopic level $\PT$-symmetric Bose-Einstein condensates have been
previously investigated with a non-Hermitian Bose-Hubbard dimer
\cite{Graefe08b, Graefe10a}.
There, gain and loss were introduced as complex on-site energy contributions.
However, the mean-field limit of such a system does not lead to the known
Gross-Pitaevskii equation with complex potentials, but instead an adapted
equation in which the nonlinear term is divided by the norm squared of the wave
function.
While this equation has the same normalized eigenstates as the Gross-Pitaevskii
equation, the dynamical behavior, including the stability properties of the
eigenstates, clearly differs \cite{Dast13b, Haag14a}.

A different approach to open quantum systems are master equations in
Lindblad form \cite{Breuer02a}, which are well established to describe phase
noise, feeding and depleting of a Bose-Einstein condensate \cite{Anglin97a,
Ruostekoski98a}.
Recently it has been shown that the mean-field limit of a master equation,
where the coherent dynamics is described by a Bose-Hubbard Hamiltonian and
single-particle losses are introduced by a Liouvillian, leads to the
Gross-Pitaevskii equation with an imaginary potential whose strength is given
by the rate of the Liouvillian \cite{Trimborn08a, Witthaut11a}.

In this paper we present a master equation describing a Bose-Einstein
condensate on two lattice sites as an open quantum system.
Gain on one lattice site and loss on the other lattice site are introduced by
two Liouvillians. 
The strengths of particle gain and loss are balanced such that it resembles the
behavior of a discrete $\PT$-symmetric Gross-Pitaevskii equation.
We show that the dynamical behavior of this master equation with balanced gain
and loss is in excellent agreement with the mean-field limit described by the
$\PT$-symmetric Gross-Pitaevskii equation.
The characteristic properties of $\PT$-symmetric systems such as the existence
of stationary states and the phase shift of the oscillations between the two
wells are also found in the many-particle description.
Visualizing the dynamics on a Bloch sphere allows us to compare the complete
dynamics of the master equation with that of the $\PT$-symmetric
Gross-Pitaevskii equation.

The remainder of this paper is ordered as follows.
In Sec.~\ref{sec:master_equation} the master equation is introduced and a
relation for the loss and gain rate is derived such that it can support
$\PT$-symmetric stationary solutions.
As shown in Sec.~\ref{sec:mean_field_limit} the mean-field limit of the master
equation leads to the $\PT$-symmetric Gross-Pitaevskii equation.
The dynamical behavior of the many-particle system is discussed in
Sec.~\ref{sec:dynamics} and compared to the mean-field limit.
Conclusions are drawn in Sec.~\ref{sec:conclusion}.

\section{Master equation with balanced gain and loss}
\label{sec:master_equation}
Ultracold atoms in an open double-well potential can be described by a quantum
master equation in Lindblad form~\cite{Anglin97a, Ruostekoski98a}.
The system considered has two discrete lattice sites with loss at site 1 and
gain at site 2 described by two Liouvillians.

The coherent dynamics is given by the Bose-Hubbard-Hamiltonian \cite{Jaksch98a,
Anglin01a} which describes bosonic atoms in the lowest-energy Bloch band of an
optical lattice,
\begin{align}
  H = &-(a_1^\dagger a_2 + a_2^\dagger a_1) \notag \\
    & + \frac{U}{2} (a_1^\dagger a_1^\dagger a_1 a_1
    + a_2^\dagger a_2^\dagger a_2 a_2),
  \label{eq:bh_hamiltonian}
\end{align}
with the bosonic creation and annihilation operators $a_j^\dagger$ and
$a_j$ acting on lattice site $j$.
The first term describes a hopping of atoms between the two lattice sites and
the second term an on-site interaction.
The strength of the on-site interaction is defined by the parameter $U$.
For comparison with the mean-field limit we introduce the macroscopic
interaction strength 
\begin{equation}
  g=(N_0-1)U,
  \label{eq:macroscopic_interaction}
\end{equation}
with the initial amount of particles in the system $N_0$.

Since the system is coupled to an environment the dynamics is governed by a
quantum master equation in Lindblad form
\begin{equation}
  \dot\rho = -\rmi [H,\rho]
    + \mathcal{L}_\mrm{loss}\rho + \mathcal{L}_\mrm{gain}\rho,
  \label{eq:master_eq}
\end{equation}
with particle loss at lattice site 1
\begin{equation}
  \mathcal{L}_\mrm{loss} \rho = -\frac12 \gamma_\mrm{loss}
    (a_1^\dagger a_1 \rho + \rho a_1^\dagger a_1
    - 2 a_1 \rho a_1^\dagger)
  \label{eq:liouvillian_loss}
\end{equation}
and particle gain at lattice site 2
\begin{equation}
  \mathcal{L}_\mrm{gain} \rho = -\frac12 \gamma_\mrm{gain}
    (a_2 a_2^\dagger \rho + \rho a_2 a_2^\dagger - 2 a_2^\dagger \rho a_2).
  \label{eq:liouvillian_gain}
\end{equation}
Localized particle loss may be induced by a focused electron
beam~\cite{Gericke08a, Wurtz09a}, whereas particle gain may be realized by
feeding from a second condensate \cite{Robins08a} using a Raman
superradiance-like pumping process \cite{Doring09a, Schneble04a, Yoshikawa04a}.

It is not clear how the ratio $\gamma_\mrm{gain}/\gamma_\mrm{loss}$ has
to be chosen such that balanced gain and loss is achieved.
We will see that the obvious choice $\gamma_\mrm{gain}=\gamma_\mrm{loss}$
is only correct in the limit $N_0\to\infty$ and a different ratio should be
chosen for a finite number of particles.

This can be understood by calculating the expectation value of the particle
number $\mean{N(t)}$ for a system consisting of only one lattice site with
either particle gain or particle loss with an initial number of particles
$N_0'$.
For this simple model we obtain analytical expressions for $\mean{N(t)}$
using the ansatz $\rho = \sum \alpha_j \ket{j}\bra{j}$, where $\ket{j}$ are the
particle number states and the coefficients $\alpha_j$ are real numbers.

In the case of particle loss the expectation value of the particle number is
given by $\mean{N_\mrm{loss}(t)} = N_0' \exp(-\gamma_\mrm{loss}t)$.
Note that this exponential decay with loss rate $\gamma_\mrm{loss}$ is exactly
the same behavior as one would obtain by introducing an imaginary potential
$V_\mrm{loss}=-\rmi \gamma_\mrm{loss}$ into the Gross-Pitaevskii equation.

In the second case of particle gain the expectation value reads
$\mean{N_\mrm{gain}(t)} = N_0' [(1+1/N_0') \exp(\gamma_\mrm{gain}t) -
1/N_0']$.
For a large number of particles $N_0'\gg1$ this leads to an exponential gain
with rate $\gamma_\mrm{gain}$, which again is exactly the same as one would
obtain by an imaginary potential $V_\mrm{gain}=\rmi \gamma_\mrm{gain}$ in the
Gross-Pitaevskii equation.

Since we want to describe the situation of balanced gain and loss the master
equation should support stationary $\PT$-symmetric solutions.
A $\PT$-symmetric state has equal probability of presence at the two lattice
sites.
Therefore we demand that if half of the particles are at the gain lattice site
and half of the particles are at the loss lattice site then, at least for short
times, the gain and loss should cancel out each other.
Expanding the terms $\mean{N_\mrm{loss}(t)}$ and $\mean{N_\mrm{gain}(t)}$ up
to the first order in $t$, introducing the total particle number at both
lattice sites $N_0=2 N_0'$ and demanding
$\mean{N_\mrm{loss}(t)}+\mean{N_\mrm{gain}(t)}=N_0$ leads to the following
condition for the gain and loss ratio
\begin{equation}
  \frac{\gamma_\mrm{gain}}{\gamma_\mrm{loss}} = 
    \frac{N_0}{N_0+2}.
  \label{eq:gainloss_ratio}
\end{equation}
This shows that $\gamma_\mrm{gain}$ has to be chosen slightly smaller than
$\gamma_\mrm{loss}$.
Only in the limit $N_0\to\infty$ the two rates have to be chosen equal.
In the following discussion gain and loss is characterized by one parameter
$\gamma=\gamma_\mrm{loss}$ and $\gamma_\mrm{gain}$ is chosen such that
Eq.~\eqref{eq:gainloss_ratio} is fulfilled.

\section{Mean-field limit}
\label{sec:mean_field_limit}
To calculate the mean-field limit of Eq.~\eqref{eq:master_eq} we follow the
procedure described in~\cite{Witthaut11a}.
There, the mean-field limit is derived for a similar system with loss but
without gain.
Starting point is the single-particle density matrix
$\sigma_{jk}=\mean{a_j^\dagger a_k}$.
The time derivative of $\sigma_{jk}$ is given by the master
equation~\eqref{eq:master_eq},
\begin{align}
  \rmi \frac{\rmd}{\rmd t} \sigma_{jk}
  = & \trace (\rmi a_j^\dagger a_k \dot{\rho})
    \notag \\
  = &-(\sigma_{j,k+1} + \sigma_{j,k-1} - \sigma_{j+1,k} - \sigma_{j-1, k})
    \notag \\
  & + U (\sigma_{kk}\sigma_{jk} - \sigma_{jj}\sigma_{jk} + \Delta_{jkkk}
    - \Delta_{jjkk})
    \notag \\
  & - \rmi \frac{\gamma_{\mrm{loss},j} + \gamma_{\mrm{loss},k}}{2} \sigma_{jk}
    \notag \\
  & + \rmi \frac{\gamma_{\mrm{gain},j} + \gamma_{\mrm{gain},k}}{2}
    (\sigma_{jk} + \delta_{jk}),
  \label{eq:spdm}
\end{align}
with the covariances 
\begin{equation}
  \Delta_{jklm}=\mean{a_j^\dagger a_k a_l^\dagger a_m}
    - \mean{a_j^\dagger a_k} \mean{a_l^\dagger a_m}
  \label{eq:covariances}
\end{equation}
and the Kronecker delta $\delta_{jk}$.
The covariances are neglected in the mean-field limit $N_0 \to \infty$
\cite{Witthaut11a}.
The difference between the terms describing gain and loss is the sign and the
additional Kronecker delta.
Due to the additional Kronecker delta the differential equation is
inhomogeneous which has the effect that there is an influx of particles from
the environment even in the case $N_0 = 0$. 
In the mean-field limit the Kronecker delta is small compared to
$\sigma_{jk}$ and can be neglected.

In our specific system we have only loss at lattice site 1 and gain at site 2,
i.e.\ $\gamma_{\mrm{loss},j}=\gamma_\mrm{loss} \delta_{1j}$ and
$\gamma_{\mrm{gain},j}=\gamma_\mrm{gain} \delta_{2j}$.
Due to Eq.~\eqref{eq:gainloss_ratio} for $N_0 \to \infty$ the two rates are
equal, $\gamma_\mrm{gain}=\gamma_\mrm{loss}=\gamma$.
In a last step the single particle density matrix is replaced by complex
amplitudes \cite{Witthaut11a}, $\sigma_{jk}=N_0 c_j^* c_k$.
With these considerations Eq.~\eqref{eq:spdm} yields the discrete
non-Hermitian Gross-Pitaevskii equation
\begin{subequations}
  \begin{align}
    \rmi \frac{\rmd}{\rmd t} c_1 = -c_2 + g |c_1|^2 c_1 
      - \rmi \frac{\gamma}{2} c_1,\\
    \rmi \frac{\rmd}{\rmd t} c_2 = -c_1 + g |c_2|^2 c_2 
      + \rmi \frac{\gamma}{2} c_2
  \end{align}
  \label{eq:discrete_gpe}%
\end{subequations}
with the macroscopic interaction strength $g$ defined in
Eq.~\eqref{eq:macroscopic_interaction}.

This shows that the gain and loss processes introduced by the Liouvillians
\eqref{eq:liouvillian_loss} and \eqref{eq:liouvillian_gain} are in the
mean-field limit described by imaginary potentials with negative and positive
sign, respectively.
The Eqs.~\eqref{eq:discrete_gpe} are evidently $\PT$-symmetric since the gain
and loss contributions have equal strength.
This system can be considered as a simple model for the more realistic extended
double-well potential with gain and loss~\cite{Dast13a, Dast13b, Haag14a}.
In fact the eigenvalue spectrum of the discrete two-mode system and the
extended double-well system are in excellent agreement~\cite{Dast13b}.

To discuss the eigenvalue spectrum the time dependence is separated
$c_j(t)=c_j\exp(-\rmi\mu t)$ leading to the time-independent Gross-Pitaevskii
equation.
The chemical potential $\mu$ can be obtained using an analytic
extension~\cite{Graefe12a},
\begin{equation*}
  \mu = 
    \begin{cases}
      \frac{g}{2} \pm \sqrt{1-\left(\frac{\gamma}{2}\right)^2}, &
        |\gamma| \leq 2,\ \PT\text{ symmetric},\\
      g \pm \rmi\gamma \sqrt{\frac14-\frac{1}{g^2+\gamma^2}}, &
        |\gamma| \geq \sqrt{4-g^2},\ \PT\text{ broken}.
    \end{cases}
    \label{eq:mf_eigenvalues}
\end{equation*}

\begin{figure}
  \centering
  \includegraphics[width=\columnwidth]{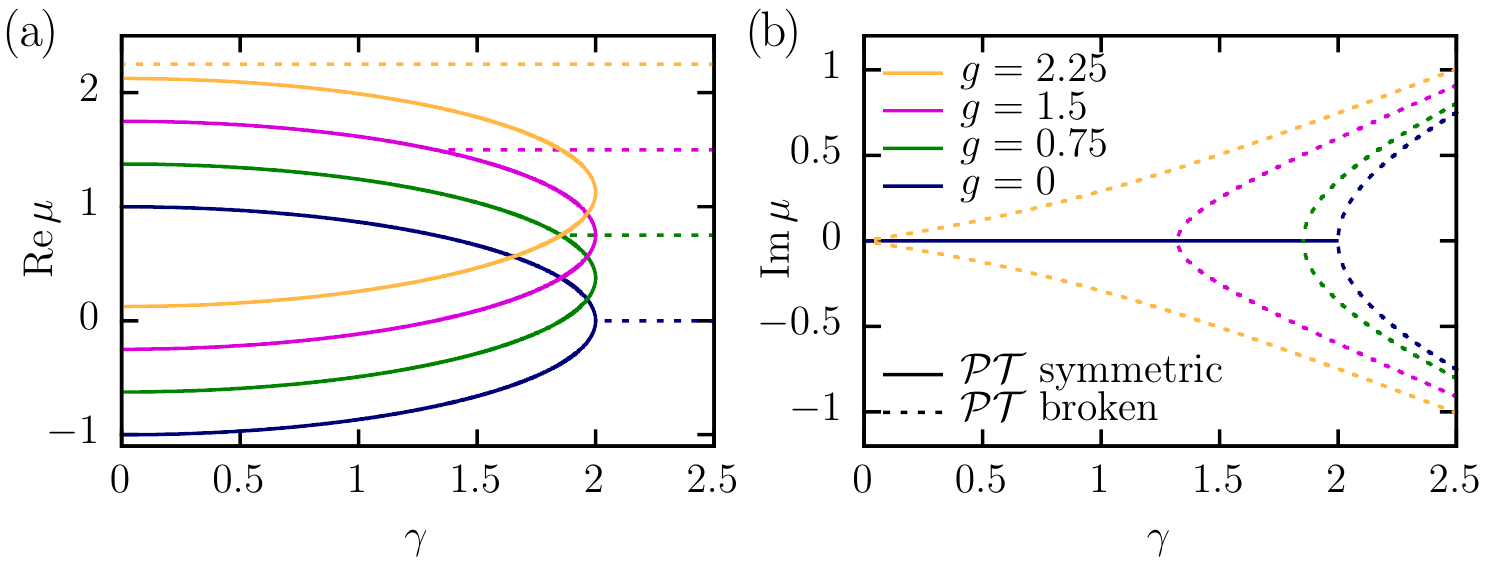}
  \caption{%
    (Color online)
    Real and imaginary parts of the eigenvalue spectrum of the $\PT$-symmetric
    discrete Gross-Pitaevskii equation \eqref{eq:discrete_gpe}.
    For $|\gamma| \leq 2$ two $\PT$-symmetric solutions with real
    eigenvalues exist.
    Two $\PT$-broken solutions with complex eigenvalues emerge at $|\gamma|
    = \sqrt{4-g^2}$.
  }%
  \label{fig:mf_spectrum}
\end{figure}

The eigenvalue spectrum is shown in Fig.~\ref{fig:mf_spectrum}.
Up to the critical value $\gamma=2$ two $\PT$-symmetric solutions with real
eigenvalues exist.
In the following we will refer to these states as the ground and the excited
state of the system.
In the linear case $g=0$ the $\PT$-broken solutions emerge from the exceptional
point at which the $\PT$-symmetric solutions vanish.
For $g>0$ the $\PT$-broken solutions emerge from the excited state and exist at
smaller values of $\gamma$.
If the nonlinearity parameter is strong enough, $g\geq2$, the $\PT$-broken
solutions exist even at $\gamma=0$.
The occurrence of symmetry-breaking states in the real potential ($\gamma=0$)
is known as macroscopic quantum self-trapping~\cite{Albiez05a}.

\section{Dynamical behavior}
\label{sec:dynamics}
If we want to compare the results of the $\PT$-symmetric Gross-Pitaevskii
equation and the master equation with balanced gain and loss  we have to
transform a mean-field state into a many-particle state.
An arbitrary mean-field state of the two-mode system is defined by two complex
numbers $\psi = (c_1,\ c_2)^T$.
In the mean-field approximation it is assumed that every particle is in the
same single-particle state.
Thus the corresponding many-particle state is $\ket{\psi} =
\prod_{j=1}^{N_0}\ket{\psi}^{(j)}$ with the single-particle state of the
$j$'th particle $\ket{\psi}^{(j)} = c_1 \ket{1}^{(j)} + c_2 \ket{2}^{(j)}$
where $\ket{1}$ and $\ket{2}$ are the states describing one particle at site 1
or 2, respectively.
Expressing $\ket{\psi}$ in the basis of Fock states with total particle number
$N_0$ leads to the result
\begin{equation}
  |\psi\rangle = \sum_{m=0}^{N_0}
    \sqrt{ \begin{pmatrix} N_0 \\ m \end{pmatrix} }
    c_1^{N_0-m}c_2^m |N_0-m,m\rangle,
  \label{eq:mf_state_fock_basis}
\end{equation}
where $|n_1, n_2 \rangle$ is a Fock state with $n_i$ particles at site $i$.
Using Eq.~\eqref{eq:mf_state_fock_basis} we can now start to compare results of
the $\PT$-symmetric Gross-Pitaevskii equation and the master equation.
The numerical results of the master equation are obtained using the quantum
jump method \cite{Plenio98a, Johansson13a} where we average over quantum
trajectories till the results converge.

\begin{figure}
  \centering
  \includegraphics[width=\columnwidth]{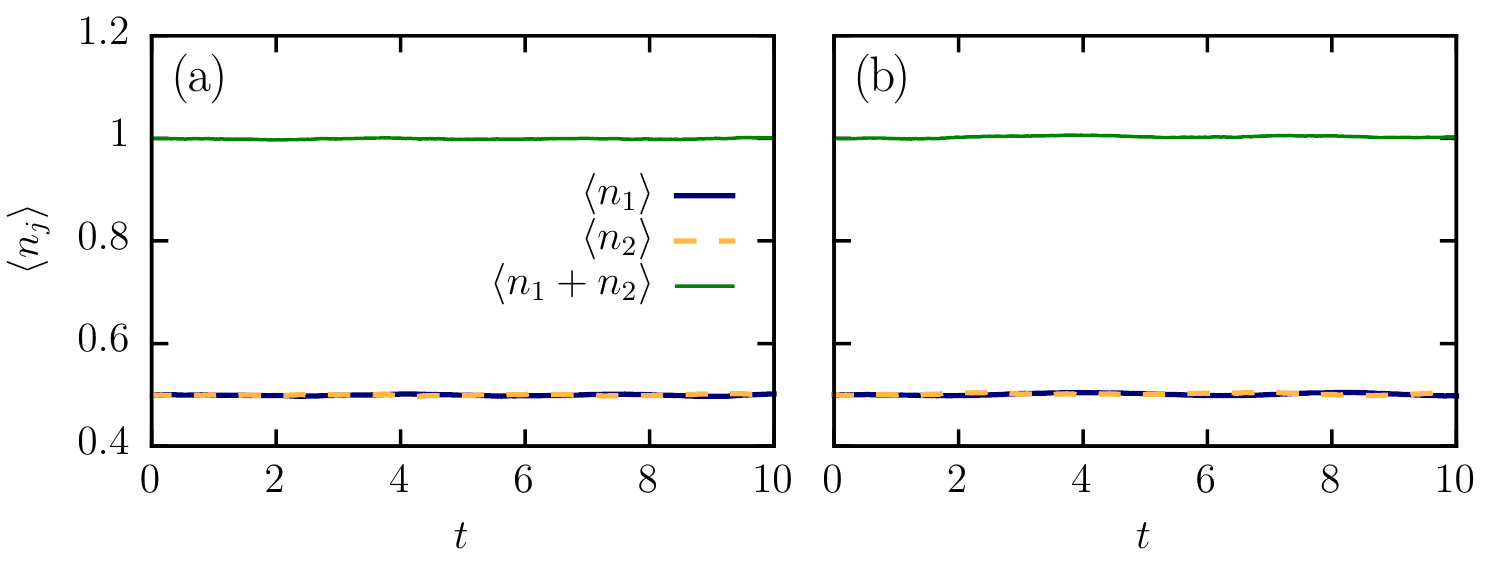}
  \caption{%
    (Color online)
    The stationary solutions of the Gross-Pitaevskii equation
    \eqref{eq:discrete_gpe} are transformed to many particle states and the
    time evolution is calculated using the master equation
    \eqref{eq:master_eq} for (a) the ground state and (b) the excited state.
    The expectation value of the particle number divided by the total initial 
    particle number at the loss site $\mean{n_1}$ and at the gain site
    $\mean{n_2}$ stay constant.
    The parameters $g=0.5$, $\gamma=0.5$ and $N_0=200$ were used and it was
    averaged over 2000 trajectories.
  }%
  \label{fig:stationary}
\end{figure}

As a first step we check if one of the most fundamental properties of
$\PT$-symmetric systems, the fact that it supports stationary solutions, is
also present in the master equation with balanced gain and loss.
Therefore we use the stationary ground state and excited state of the
$\PT$-symmetric discrete Gross-Pitaevskii equation~\eqref{eq:discrete_gpe},
transform the mean-field state into a many-particle state using
Eq.~\eqref{eq:mf_state_fock_basis} and calculate the time evolution of this
state with the master equation~\eqref{eq:master_eq}.
The result is shown in Fig.~\ref{fig:stationary}
for both the stationary ground state and the excited state.
This shows that the stationary solutions of the $\PT$-symmetric
Gross-Pitaevskii equation can be transfered to the master equation with
balanced gain and loss, and again behave stationary in the sense that the
expectation values of the particle number at both lattice sites are constant.
Thus this fundamental property of $\PT$-symmetric systems is also present in
the master equation.
Note that these are not steady states which satisfy $\dot{\rho}=0$.

\begin{figure}
  \centering
  \includegraphics[width=\columnwidth]{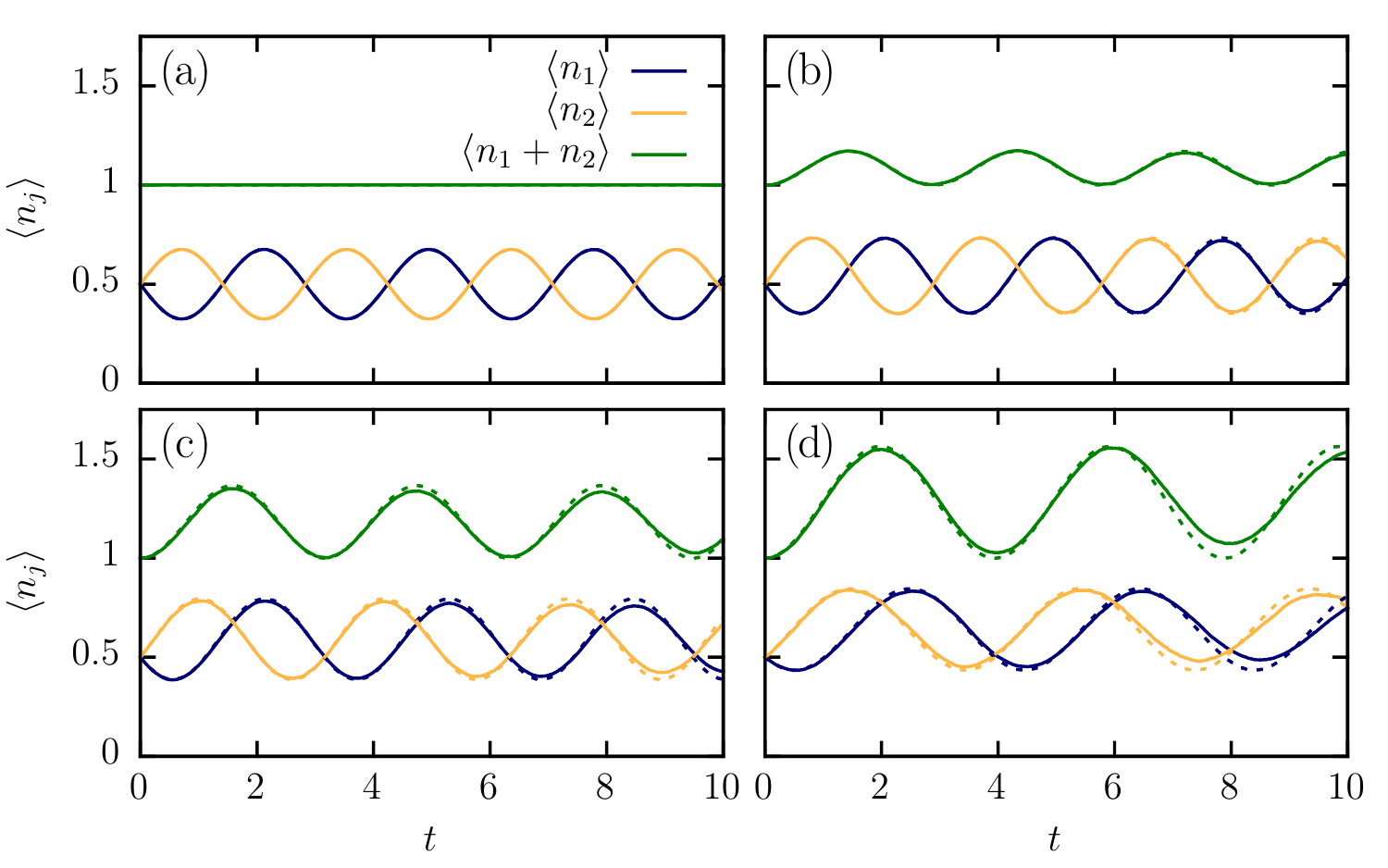}
  \caption{%
    (Color online)
    The expectation value of the particle number at the loss site $\mean{n_1}$,
    the gain site $\mean{n_2}$, and at both sites divided by the initial amount
    of particles in the system $N_0=100$ is shown for (a) $\gamma=0$, (b)
    $\gamma=0.5$, (c) $\gamma=1$ and (d) $\gamma=1.5$.
    The initial wave functions are superpositions of the stationary states
    \eqref{eq:wf_superpos} with $\theta=0.2$.
    The strength of the on-site interaction is $g=0.5$ and it was averaged over
    500 trajectories.
    The oscillations at the two lattice sites become more and more in phase as
    $\gamma$ is increased.
    The calculations using the master equation (solid lines) are in excellent
    agreement with the results of the $\PT$-symmetric Gross-Pitaevskii equation
    (dashed lines).
    The dashed lines are exactly on top of the solid lines in (a) and (b).
    Small deviations can only be seen in (c) and (d) for large times.
  }%
  \label{fig:pulsing_U_0k5}
\end{figure}

As a next step we want to investigate not only stationary solutions but
oscillations between the two lattice sites.
Fig.~\ref{fig:pulsing_U_0k5} shows the time evolution of the expectation
value of the particle number at the gain site, the loss site and the total
particle number for different values of the gain-loss parameter $\gamma$.
The initial wave functions are superpositions
\begin{equation}
  \ket{\psi}=\cos\theta \ket{\psi_g} + \sin\theta \ket{\psi_e}
  \label{eq:wf_superpos}
\end{equation}
of the stationary ground state $\ket{\psi_g}$ and excited state $\ket{\psi_e}$
which fulfill exact $\PT$symmetry, $\PT \ket{\psi_{g/e}} = \ket{\psi_{g/e}}$.

For $\gamma=0$ the dynamics is coherent and thus the total amount of particles
in the system stays constant.
The oscillations at the two lattice sites have a phase difference of $\pi$,
thus the maxima and minima coincide.
If gain and loss are introduced into the system the dynamics is no longer
coherent and as a result the total amount of particles oscillates.
The oscillation of the total amount of particles becomes stronger for greater
values of $\gamma$.
The reason for this behavior is that the oscillations at the lattice sites
become more and more in phase as $\gamma$ increases and the exceptional point
at $\gamma=2$ is approached (see Fig.~\ref{fig:mf_spectrum}).
This behavior is characteristic of $\PT$-symmetric systems and has already
been discussed for Bose-Einstein condensates in a spatially extended
potential~\cite{Cartarius12b, Dast13a}, and was experimentally confirmed in
optical systems~\cite{Klaiman08a, Ruter10a}.

\begin{figure}
  \centering
  \includegraphics[width=\columnwidth]{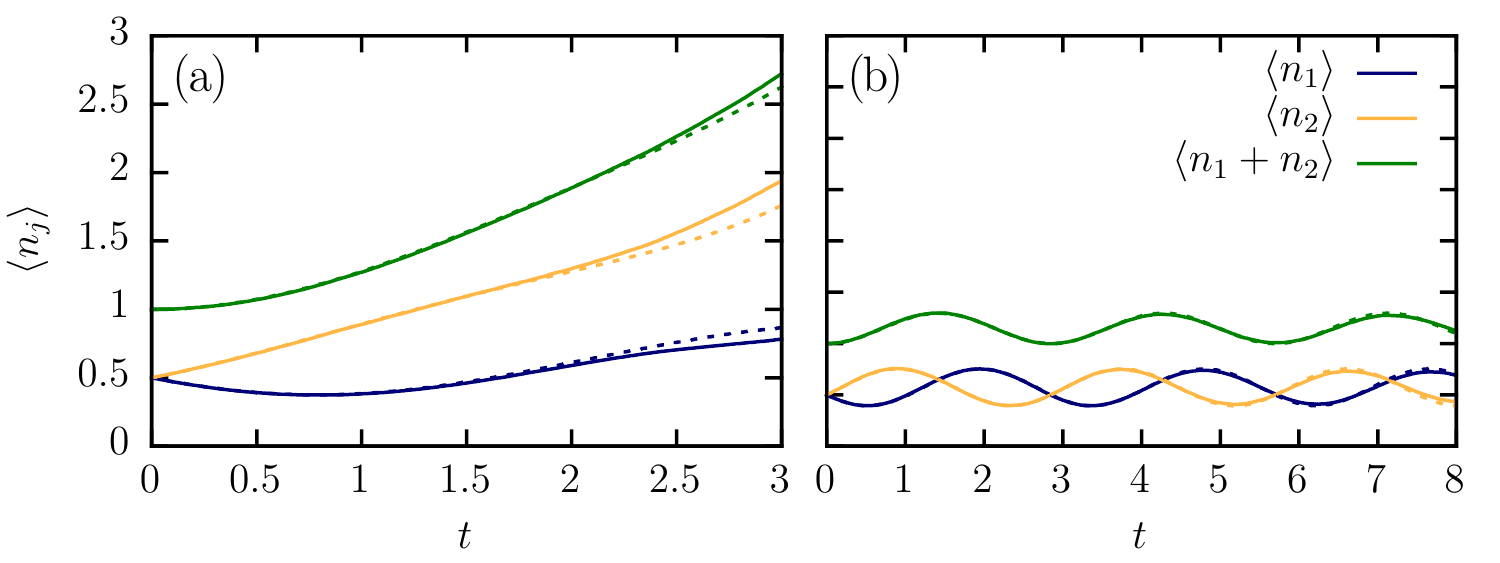}
  \caption{%
    (Color online)
    The expectation value of the particle number for two different initial wave
    functions.
    The initial wave functions are superpositions of the stationary states
    \eqref{eq:wf_superpos} with (a) $\theta = 1.4$ and (b) $\theta = 0.2$.
    The parameters $g=1$, $\gamma=1$, $N_0=100$ are used and the expectation
    values were averaged over 500 trajectories.
    Depending on the initial superposition the number of particles (a) diverges
    or (b) oscillates.
    Again the results of the master equation (solid lines) and the
    Gross-Pitaevskii equation (dashed lines) are in excellent agreement.
    In (b) the dashed lines are not even visible since they lie exactly on top
    of the solid lines.
  }%
  \label{fig:explosion}
\end{figure}

Since the system considered is nonlinear it is possible that for the same
system parameters one superposition of the ground state and the excited states
shows stable oscillations while another superposition diverges.
Such an explosion of the condensate's number of particles has been discussed in
\cite{Cartarius12a, Dast13a} for an extended potential and a double-$\delta$
potential, respectively.
The same behavior is also found using the master equation with balanced gain
and loss as shown in Fig.~\ref{fig:explosion}.

Both Fig.~\ref{fig:pulsing_U_0k5} and Fig.~\ref{fig:explosion} show the
mean-field dynamics of the Gross-Pitaevskii equation in comparison to the
many-particle dynamics of the master equation.
The dynamics are in excellent agreement and only for strong values of the
gain-loss parameter $\gamma$ or long times deviations are observable.

\begin{figure}
  \centering
  \includegraphics[width=0.7\columnwidth]{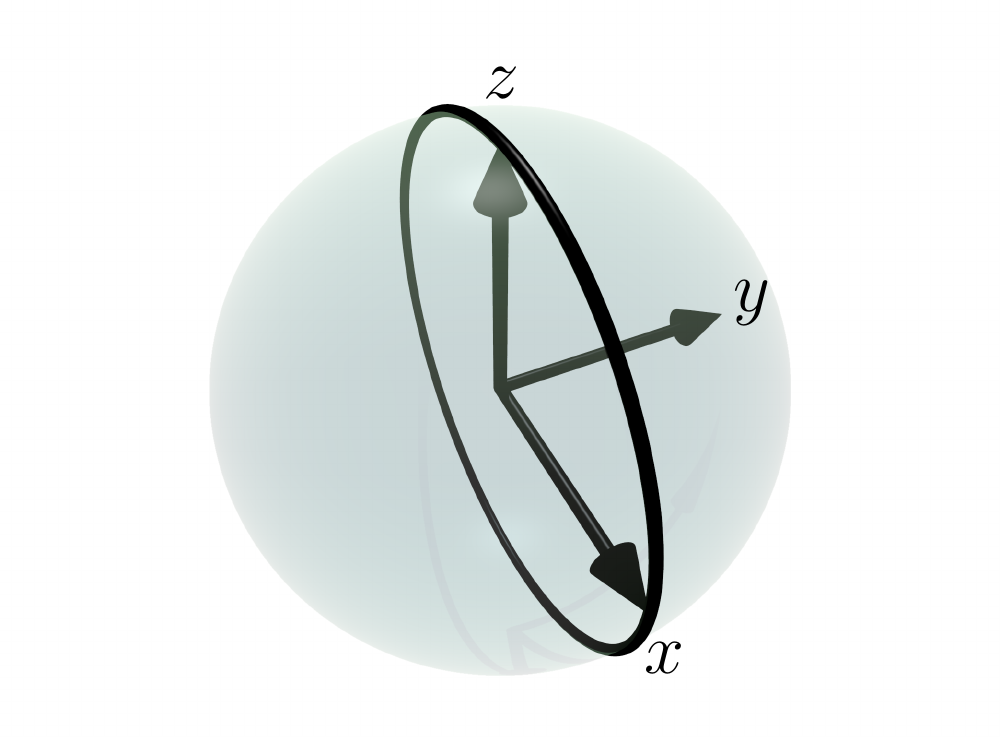}
  \caption{%
    The coordinate system used for the Bloch vector $b_\alpha =
    \mean{\Sigma_\alpha}$, $\alpha=x,y,z$.
    The north pole corresponds to the stationary excited state $\ket{\psi_e}$
    of the system in the mean-field limit and the south pole is the state
    orthogonal to $\ket{\psi_e}$ in the two-dimensional space spanned by
    $\ket{\psi_e}$ and the stationary ground state.
    In the Hermitian case the south pole represents exactly the ground state.
    All initial states reside on the great circle in the $xz$-plane.
  }%
  \label{fig:bloch_basis}
\end{figure}

The previous calculations showed that fundamental properties of $\PT$-symmetric
systems are also found in the many-particle system described by the master
equation with balanced gain and loss.
However, the time evolution was only discussed for a few wave packets
as initial wave functions.
To gain a complete picture of the dynamical behavior the visualization on a
Bloch sphere has already proved to be useful for $\PT$-symmetric
systems~\cite{Graefe08b, Haag14a}.
To map the dynamics onto the Bloch sphere we define the many-particle operator
\begin{equation}
  \Sigma_\alpha = \sum_{j=1}^{N} \sigma_{\alpha,j}, \quad \alpha = x,y,z,
  \label{eq:sigma_manyparticle}
\end{equation}
with the Pauli matrices $\sigma_{\alpha,j}$ acting on the $j$'th particle.
The Bloch vector $\mbf{b}$ is defined by the expectation value of this
operator, $b_\alpha = \mean{\Sigma_\alpha}$ and is plotted using the coordinate
system shown in Fig.~\ref{fig:bloch_basis}.
In second quantization Eq.~\eqref{eq:sigma_manyparticle} reads
\begin{equation}
  \Sigma_\alpha = \sum_{i,j=1}^{2} \braopket{i}{\sigma_\alpha}{j}
    a_i^\dagger a_j, \quad \alpha = x,y,z,
  \label{eq:sigma_manyparticle_sq}
\end{equation}
where $\ket{i} \in \{\ket{1}, \ket{2}\}$ are, as before, the one-particle
states describing a particle at lattice site 1 or 2, respectively.

The Pauli matrices are defined in the basis of the Bloch sphere $\{\ket{e_1},
\ket{e_2}\}$
\begin{subequations}
  \begin{align}
    \sigma_x &= \ket{e_1}\bra{e_2} + \ket{e_2}\bra{e_1},\\
    \sigma_y &= -\mrm{i}\ket{e_1}\bra{e_2} + \mrm{i}\ket{e_2}\bra{e_1},\\
    \sigma_z &= \ket{e_1}\bra{e_1} - \ket{e_2}\bra{e_2}.
  \end{align}
  \label{eq:spin_matrices}%
\end{subequations}
The first basis vector of the Bloch sphere points to the north pole and is
chosen to be the stationary excited state of the system,
\begin{equation}
  \ket{e_1} = \ket{\psi_e} = c_1 \ket{1} + c_2 \ket{2}.
  \label{eq:bloch_e1}
\end{equation}
The second basis vector pointing to the south pole of the Bloch sphere is
orthogonal to the first basis vector 
\begin{equation}
  \ket{e_2} = \mrm{i} (-c_2^* \ket{1} + c_1^* \ket{2} ),
  \label{eq:bloch_e2}
\end{equation}
and the phase is chosen such that it is exactly $\PT$ symmetric.
Note that only in the Hermitian case $\ket{e_2}$ is equal to the stationary
ground state.

Using the Eqs.~\eqref{eq:spin_matrices}--\eqref{eq:bloch_e2} allows us to
calculate the coefficients of the operator in
Eq.~\eqref{eq:sigma_manyparticle_sq},
\begin{subequations}
  \begin{align}
    \sigma_x &=
    \begin{pmatrix}
      -2 \imag(c_1 c_2) & -\rmi (c_1^2 + (c_2^*)^2) \\
      \rmi ((c_1^*)^2 + c_2^2) & 2 \imag(c_1 c_2)
    \end{pmatrix},
    \\
    \sigma_y &=
    \begin{pmatrix}
      2 \real(c_1 c_2) & -c_1^2 + (c_2^*)^2 \\
      -(c_1^*)^2 + c_2^2 & -2 \real(c_1 c_2)
    \end{pmatrix},
    \\
    \sigma_z &=
    \begin{pmatrix}
      |c_1|^2 - |c_2|^2 & 2 c_1 c_2^* \\
      2 c_1^* c_2 & |c_2|^2 - |c_1|^2
    \end{pmatrix}.
  \end{align}
  \label{eq:sigma_alpha}%
\end{subequations}
\begin{figure*}
  \centering
  \includegraphics[width=\textwidth]{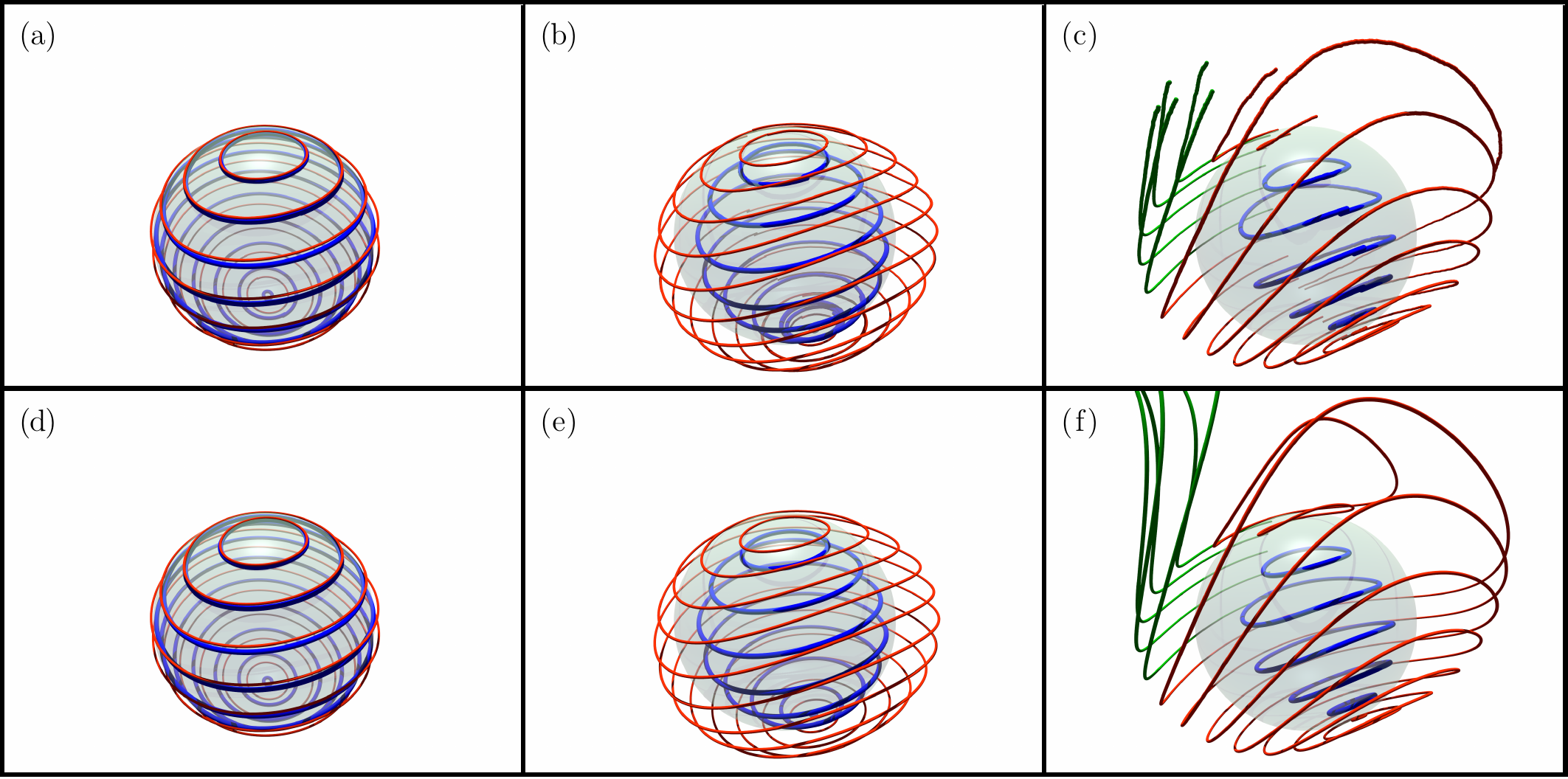}
  \caption{%
    (Color online)
    Dynamics on a Bloch sphere described by the master equation with balanced
    gain and loss (upper panels) and the $\PT$-symmetric Gross-Pitaevskii
    equation (lower panels), respectively.
    In all graphs the coordinate system introduced in
    Fig.~\ref{fig:bloch_basis} was used and all spheres are aligned
    appropriately.
    The gain-loss parameter is $\gamma=0.1$ in the left panels, $\gamma=0.7$ in
    the middle panels and $\gamma=1.3$ in the right panels.
    The parameters $g=0.5$, $N_0=50$ (a,b), $N_0=100$ (c) were used and it was
    averaged over 500 trajectories.
    The elliptic fixed point on the north pole is the excited state of the
    system.
    The ground state of the system is the second fixed point which for
    $\gamma=0$ resides on the south pole and wanders towards the north pole as
    $\gamma$ increases.
    The many-particle calculations and the mean-field calculations are in
    excellent agreement.
  }%
  \label{fig:bloch_sphere}
\end{figure*}
Since the system is coupled to an environment initial pure states become
statistical mixtures.
For pure states the norm of the Bloch vector is equal to the amount of
particles in the system.
The norm of the Bloch vector of statistical mixtures, however, is smaller than
the amount of particles in the system and, as a result, such states reside in
the interior of the Bloch sphere.
Since the number of particles is not constant both effects, the in/outflux of
particles and the decoherence, lead to a change in the norm of the Bloch
vector.
To separate these effects the Bloch vector is always normalized to the
expectation value of the particle number.
This allows us to directly compare the many-particle dynamics with that of the
mean-field description given by the $\PT$-symmetric Gross-Pitaevskii equation
which only can cover pure states.

The dynamics on the Bloch sphere is shown in Fig.~\ref{fig:bloch_sphere}.
The calculations using the master equation with balanced gain and loss (upper
panels) are compared with the dynamics of the $\PT$-symmetric Gross-Pitaevskii
equation (lower panels).
All initial states are normalized pure states and are chosen such that
they start on a great circle of the Bloch sphere through the north pole, the
south pole and the ground state of the system (see Fig.~\ref{fig:bloch_basis}).
These initial states are $\PT$ symmetric since all states in the $xz$-plane
fulfill this symmetry \cite{Haag14a}.

Fig.~\ref{fig:bloch_sphere}(a) shows the dynamics for $\gamma=0.1$.
There are two elliptic fixed points, the excited state on the north pole and
the ground state which is almost at the south pole.
Only for $\gamma=0$ the ground state resides on the south pole because in this
case the two stationary states are orthogonal.
Due to the coupling to the environment the particle number is not conserved and
thus the trajectories do not run on the surface of the Bloch sphere.
The sum of the trajectories defines two distinct closed surfaces, one inside
the Bloch sphere (thick blue lines) and one outside (red lines), thus
describing oscillations to fewer or more particles than the original amount,
respectively.
These closed surfaces cannot be penetrated by other trajectories.

Increasing the gain-loss parameter to $\gamma=0.7$ leads to the dynamics shown
in Fig.~\ref{fig:bloch_sphere}(b).
As $\gamma$ is increased the ground state wanders towards the north pole on the
front side of a great circle through the two poles.
Due to the stronger coupling to the environment more particles are exchanged
and the trajectories depart further off the Bloch sphere.
Again we recognize the two distinct closed surfaces inside and outside of the
sphere.

The Bloch sphere for $\gamma=1.3$ in Fig.~\ref{fig:bloch_sphere}(c) shows an
additional type of trajectories (green lines).
The trajectories outside the sphere no longer define a closed surface.
Some of the trajectories are still periodic (red lines) while other
trajectories diverge to higher radii (green lines).
The diverging trajectories are guided by the $\PT$-broken eigenstates of the
system as discussed in~\cite{Haag14a}.

The lower three panels of Fig.~\ref{fig:bloch_sphere} show the dynamics
described by the $\PT$-symmetric Gross-Pitaevskii equation for comparison.
For $\gamma=0.1$ and $\gamma=0.7$ the mean-field dynamics and the many-particle
dynamics are in excellent agreement.
For $\gamma=1.3$ the agreement is again very good, solely the trajectories at
large radii are cut off in the many-particle calculations.
The reason for this behavior is that the maximum amount of particles in the
system is limited by the choice of the Fock basis.

The comparison shows that although a relatively small particle number of
50--100 was used for the many-particle calculations an excellent agreement with
the $\PT$-symmetric Gross-Pitaevskii equation is found.

\section{Conclusion}
\label{sec:conclusion}
We have investigated an open quantum system described by a master equation
\eqref{eq:master_eq} in Lindblad form whose mean-field limit is a
$\PT$-symmetric Gross-Pitaevskii equation \eqref{eq:discrete_gpe}.
The numerical treatment has shown that the characteristic properties known
from nonlinear $\PT$-symmetric systems are also found in the many-particle
dynamics described by the master equation with balanced gain and loss.

In particular we showed that the stationary solutions of the $\PT$-symmetric
Gross-Pitaevskii equation behave also stationary in the many-particle
description using the master equation with balanced gain and loss.
The master equation supports characteristic dynamical properties of
$\PT$-symmetric systems such as the in-phase pulsing between the lattice sites
if the gain and loss is increased.
The comparison using the Bloch sphere formalism goes one step further since it
characterizes the whole dynamics of the system including the stability
properties.
Since the Bloch sphere behavior showed an excellent agreement we can conclude
that the master equation with balanced gain and loss is indeed the adequate
many-particle description of a $\PT$-symmetric Bose-Einstein condensate.
This supports the usual interpretation that the imaginary potentials introduced
for the Gross-Pitaevskii equation model an in- or outflux of atoms coherently
coupled to the condensate.

These results are a step towards a microscopic understanding of $\PT$-symmetric
quantum systems and opens the way to investigate many-particle effects such
as correlations which are not accessible in the mean-field description.

\end{document}